\input{aipcheck}

\documentclass[
    ,final            
  ]
  {aipproc}

\layoutstyle{8x11single}

\begin{document}

\title{An on-line library of afterglow light curves}


\author{Hendrik J. van Eerten, Andrew I. MacFadyen and Weiqun Zhang}{
  address={Center for Cosmology and Particle Physics, Physics Department, New York University, New York, NY 10003}
}




\begin{abstract}
Numerical studies of afterglow jets reveal significant qualitative differences with simplified analytical models. We present an on-line library of synthetic afterglow light curves and broadband spectra for use in interpreting observational data. Light curves have been calculated for various physics settings such as explosion energy and circumburst structure, as well as differing jet parameters and observer angle and redshift. Calculations gave been done for observer frequencies ranging from low radio to X-ray and for observer times from hours to decades after the burst. The light curves have been calculated from high-resolution 2D hydrodynamical simulations performed with the RAM adaptive-mesh refinement code and a detailed synchrotron radiation code. 

The library will contain both generic afterglow simulations as well as specific case studies and will be freely accessible at \url{http://cosmo.nyu.edu/afterglowlibrary}. The synthetic light curves can be used as a check on the accuracy of physical parameters derived from analytical model fits to afterglow data, to quantitatively explore the consequences of varying parameters such as observer angle and for accurate predictions of future telescope data.
\end{abstract}

\maketitle


\section{Introduction}

Gamma-ray burst (GRB) afterglows are well described by radiation from outflows expanding into an external medium \cite{Zhang2004}. At first the outflow takes the shape of two highly collimated relativistic jets ejected in opposite directions. Over time, the jets slow down and gradually widen, eventually resulting in a spherical nonrelativistic blast wave. Synchrotron radiation is produced by shock-accelerated electrons interacting with a shock-generated magnetic field. The radiation will peak at progressively longer wavelengths and the observed light curve will change shape whenever the observed frequency crosses into different spectral regimes as determined by the velocity of the flow and the physics of the radiative process, such as synchrotron self-absorption and electron cooling.

Analytical models have greatly enhanced our understanding of GRB afterglows, but are limited in that they rely on simplifications of the fluid properties and radiation mechanisms involved. For example, the flux received by observers not on the jet axis has usually been calculated assuming that all emission originates from a homogeneous slab at the shock front, rather than taking into account the full fluid profile \cite{Kumar2000, Granot2002}. Also, lateral spreading of the jet is often ignored completely or roughly estimated assuming angle-independent values for the fluid variables at a shock front that expands laterally at the speed of sound \cite{Rhoads1999}. 

High resolution relativistic hydrodynamical jet simulations help to overcome the limitations of analytical models. The aim of the current work is to present the results of jet simulations performed using the \textsc{RAM} adaptive-mesh refinement code \cite{Zhang2006}, linked to a synchrotron radiation module \cite{Zhang2009, vanEerten2009, vanEerten2010}. Synthetic light curves are made available to the community and are freely accessible at \url{http://cosmo.nyu.edu/afterglowlibrary}. The synthetic light curves can be used as a check on the accuracy of physical parameters derived from analytical model fits to afterglow data, to quantitatively explore the consequences of varying parameters such as observer angle and for accurate predictions of future telescope data.

\section{Jet dynamics}

\begin{figure}
  \includegraphics[width=0.9\columnwidth]{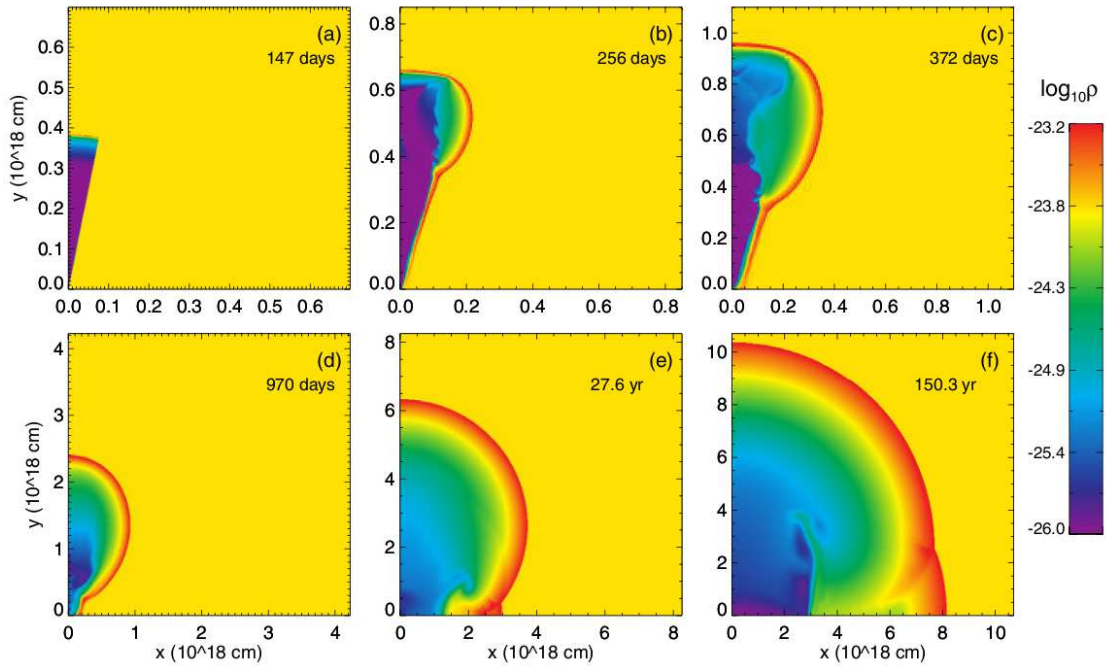}
  \caption{Time evolution of the comoving density for a typical GRB. The density (g cm$^{-3}$) is color coded on a logarithmic scale. Snapshots of the simulation are shown at (a) 147 days, (b) 256 days, (c) 372 days, (d) 970 days, (e) 27.6 years and (f) 150.3 years in the lab frame time after the explosion. The beginning of the simulation is at 147 days. Panel (d) shows that the GRB outflow is still highly anisotropic at $t \approx 970$ days, where analytical estimates have put the transition to nonrelativistic flow \cite{Piran2005}. This figure has been taken from \cite{Zhang2009}, where a more detailed discussion of the simulation can be found as well.}
  \label{dynamics_figure}
\end{figure}

We use the \textsc{RAM} adaptive mesh hydrodynamics code \cite{Zhang2006} to calculate the jet evolution in 2D. We start with a conic section of the self-similar Blandford-McKee analytic solution for a relativistic explosion \cite{Blandford1976}. The adaptive mesh approach allows us to dynamically refine the grid where necessary, leading to an enormous increase in effective resolution. The jet evolution is followed for a long time, until the explosion has become nearly spherical due to lateral spreading of the jet and until the shock velocity has become nonrelativistic. This stage is again self-similar and can be described by the Sedov-Taylor solution. Fig. \ref{dynamics_figure} shows the comoving density for a typical jet (half opening angle of 0.2 rad, total energy of $2\times10^{51}$ erg in both jets and homogeneous circumburst number density of 1 cm$^{-3}$) at various stages.

The jet energy, opening angle and circumburst density profile vary between simulations. The maximum refinement level is set very high at first (the precise level being chosen is such that the initial Blandford-McKee profile is properly resolved). Over time the maximum refinement level is decreased gradually, following the widening of the shock profile. In order to keep the calculation time for a single simulation manageable, we also keep at low refinement level the region deep within the shock. This region does not contribute to the outgoing emission or overall jet dynamics. But if it is not articially derefined, the Kelvin-Helmholtz instabilities that develop in the region will slow down the computation time significantly. For similar reasons we keep  at a low refinement level the nonrelativistic shock that initially develops between the empty cavity far behind the forward shock and the circumburst medium.

Simulations have found that very little sideways expansion takes place for the ultrarelativistic material near the forward shock, while the mildly relativistic and Newtonian jet material further downstream undergoes more sideways expansion. When taking a fixed fraction of the total energy contained within an opening angle as a measure of the jet collimation it is found that sideways expansion is logarithmic (and not exponential, as used by some early analytic models such as that of \cite{Rhoads1999}). This sideways expansion sets in when the Lorentz factor of the jet is approximately the inverse of the original jet half opening angle. The jet becomes nonrelativistic when the isotropic equivalent energy in the jet becomes approximately equal to the rest mass energy of the material swept up by a spherical explosion. The transition to spherical flow was found to be a slow process and was found to take five times longer than the transition to nonrelativistic flow. After this time the outflow can be described by the Newtonian Sedov-von Neumann-Taylor solution.

\section{Synchrotron radiation}

The dominant emission mechanism in the afterglow phase is synchrotron radiation. We calculate the radiation from a simulation output using one of two different methods, depending on whether synchrotron self-absorption plays a role at the observer frequencies of interest. Both approaches have in common their parametrization of shock acceleration of electrons and the subsequent evolution of the electron distribution. The energy density in the accelerated electrons is assumed to be a fraction $\epsilon_E$ (typically $\sim 0.1$) of the thermal energy density, the magnetic field energy density a fraction $\epsilon_B$ (typically $\sim 0.01$) of the thermal energy density and the number density of accelerated electrons a fraction $\xi_N$ (typically $\sim 1.0$) of the fluid number density. A power law in energy with lower limit $\gamma_m$ and with slope $p$ (typically $\sim -2.5$) is assumed for the accelerated electron distribution. The local synchrotron peak frequency $\nu_m$ is calculated from $\gamma_m$, which in turn is fixed by $\epsilon_E$, $\epsilon_B$ and $\xi_N$ and standard synchrotron theory. This approach follows \cite{Sari1998}.

If self-absorption can be ignored, the emission is calculated following the approach taken in \cite{Zhang2009, vanEerten2010c} (these papers also present an analysis of light curves obtained by this method). The procedure is as follows. For every grid cell in the simulation, the emission is calculated (taking into account the beaming factor, observer redshift etc.) and added to the appropiate observer time bin. Electron cooling is treated by assuming a global cooling time that is equal to the time since  the explosion.

If self-absorption does play a role, rays of emission through the expanding fluid need to be calculated explicitly by simultaneously solving a large set of linear radiative transfer equations. The number of rays that are followed through the changing fluid is adapted dynamically depending on the local changes in intensities via a procedure analogous to adaptive mesh refinement for the fluid dynamics calculation. Light curves obtained using this method have been published in \cite{vanEerten2010, vanEerten2010b}. Additional advantages of this method are that it is easy to determine the exact regions of the fluid that contribute to the emerging flux and that spatially resolved afterglow images are automatically generated as a by-product of the flux calculation.

\section{Afterglow Database}

\begin{figure}
  \includegraphics[width=0.9\columnwidth]{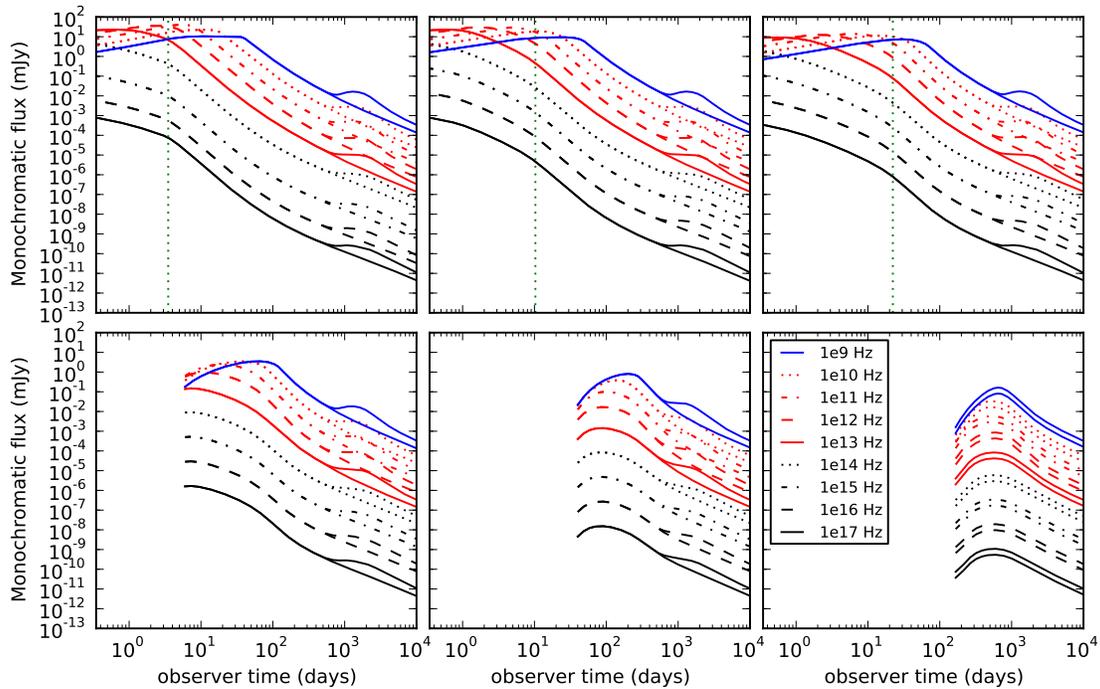}
  \caption{Simulated light curves for various observer angles and frequencies for a jet with half opening angle of 0.2 rad, total energy of $2\times10^{51}$ erg in both jets and homogeneous circumburst number density of 1 cm$^{-3}$. On the top row we have small observer angles, 0.0, 0.1 and 0.2 radians from left to right. On the bottom row we have large observer angles, 0.4, 0.8 and 1.57 radians from left to right. Observer frequencies from $10^9$ Hz up to $10^{17}$ Hz are plotted. For each frequency and angle a curve is plotted both with and without the contribution from the counterjet. The vertical lines in the top plots indicate jet break time estimates. The legend in the bottom right plot refers to all plots. Large observer angles are truncated at earlier time to show only observation times completely covered by the simulation. The figure has been taken from \cite{vanEerten2010c}, where the dataset and its features such as the jet break times are discussed in more detail.}
  \label{light_curves_figure}
\end{figure}

The on-line database at \url{http://cosmo.nyu.edu/afterglowlibrary} will aim to contain all light curves that have been calculated using \textsc{ram} and the synchrotron radiation code. An example set of light curves from \cite{vanEerten2010c} is shown in Fig. \ref{light_curves_figure}. Light curve datasets are available in different formats, including plain text. An overview plot is provided which each afterglow dataset, showing a pre-selection of light curves from some of the available frequencies. The rest of the data can be downloaded directly and additional scripts are provided to quickly visualize the dataset.

At the moment, an afterglow dataset can be selected either by selecting a specific GRB from the list or by selecting physical explosion parameters from a list. Such parameters include observer redshift, angle, explosion energy, circumburst density as well as radiation parameters $p$, $\epsilon_E$, $\epsilon_B$, $\xi_N$. Frequencies are chosen to match commonly used observer frequencies, like in \cite{vanEerten2010c} and in our second contribution presented elsewhere in these proceedings\footnote{see \emph{Off-Axis Afterglow Light Curves from High-Resolution Hydrodynamical Jet Simulations}, H.J. van Eerten, A.I. MacFadyen \& W. Zhang, elsewhere in these proceedings.}, where observed X-ray flux is calculated at 1.5 keV (observable by e.g. \emph{Swift}) and radio flux at 8.46 Ghz (e.g \textsc{VLA, WSRT}). Alternatively and depending on the aim of the publication that first presents the light curves, a whole range of frequencies between $10^9$ Hz (or lower, $10^6$ Hz) and $10^{17}$ Hz is occasionally used to show the complete broadband picture (see also \cite{vanEerten2010c} and Fig. \ref{light_curves_figure}).

The downloadable synthetic light curves can be used as a check on the accuracy of physical parameters derived from analytical model fits to afterglow data and as an aid in case studies. By comparing different light curves when a single input parameter is varied (e.g. observer angle or jet opening angle) it is possible to quickly and quantitatively explore the impact of these paremeters on the observed signal. Finally, various new telescopes, like \textsc{lofar} and \textsc{ska} are under development that will probe new parts of the spectrum and synthetic light curves can be used as a basis for predicting what these telescopes will observe from GRB afterglows.

This paper broadly follows \cite{Zhang2009, vanEerten2010c} in its presentation and discusses the same dataset. 


\begin{theacknowledgments}This work was supported in part by NASA under Grant
No. 09-ATP09-0190 issued through the Astrophysics Theory Program
(ATP).  The software used in this work was in part developed by the
DOE-supported ASCI/Alliance Center for Astrophysical Thermonuclear
Flashes at the University of Chicago.
\end{theacknowledgments}



\bibliographystyle{aipproc}   

\bibliography{afterglowlibrary}

\IfFileExists{\jobname.bbl}{}
 {\typeout{}
  \typeout{******************************************}
  \typeout{** Please run "bibtex \jobname" to optain}
  \typeout{** the bibliography and then re-run LaTeX}
  \typeout{** twice to fix the references!}
  \typeout{******************************************}
  \typeout{}
 }

\end{document}